\documentstyle[amssymb,floats,aps,prb,epsf]{revtex}
\epsfclipon 
\begin{document}
\twocolumn[\hsize\textwidth\columnwidth\hsize\csname
@twocolumnfalse\endcsname 

\draft
\title{Charge transport in underdoped bilayer cuprates}

\author{Feng Yuan}
\address{Department of Physics, Beijing Normal University, Beijing
100875, China\\
The Abdus Salam International Centre for Theoretical Physics, 34014
Trieste, Itlay}

\author{Jihong Qin and Shiping Feng}
\address{Department of Physics and Key Laboratory of Beam
Technology and Material Modification, Beijing Normal University,
Beijing 100875, China\\
Interdisciplinary Center of Theoretical Studies, Chinese Academy
of Sciences, Beijing 100080, China\\
National Laboratory of Superconductivity, Chinese Academy of
Sciences, Beijing 100080, China}

\author{Wei Yeu Chen}
\address{Department of Physics, Tamkang University, Tamsui 25137,
Taiwan}
\date{Received 13 November 2002}

\maketitle

\begin{abstract}
Within the $t$-$J$ model, we study the charge transport in
underdoped bilayer cuprates by considering the bilayer interaction.
Although the bilayer interaction leads to the band splitting in
the electronic structure, the qualitative behavior of the charge
transport is the same as in the case of single layer cuprates.
The conductivity spectrum shows a low-energy peak and the unusual
midinfrared band. This midinfrared band is suppressed severely with
increasing temperatures, while the resistivity in the heavily
underdoped regime is characterized by a crossover from the high
temperature metallic-like to the low temperature insulating-like
behaviors, which are consistent with the experiments.
\end{abstract}
]
\bigskip
\narrowtext

It has become clear in the past ten years that cuprate
superconductors are among the most complex systems studied in
condensed matter physics \cite{n1,n2}. The complications arise
mainly from (1) strong anisotropy in the properties parallel and
perpendicular to the CuO$_{2}$ planes which are the key structural
element in the whole cuprate superconductors, and (2) extreme
sensitivity of the properties to the compositions (stoichiometry)
which control the carrier density in the CuO$_{2}$ plane, and
therefore the regimes have been classified into the underdoped,
optimally doped, and overdoped, respectively \cite{n1,n2}. In the
underdoped and optimally doped regimes, the experimental results
\cite{n3} show that the ratio of the c-axis and in-plane
resistivities $R=\rho_{c}(T)/\rho_{ab}(T)$ ranges from $R\sim 100$
to $R>10^{5}$, which reflects that the charged carriers are tightly
confined to the CuO$_{2}$ planes. This large magnitude of the
resistivity anisotropy also leads to the general notion that the
physics of doped cuprates is almost entirely two-dimensional, and
can be well described by a single CuO$_{2}$ plane \cite{n4}.
However, this picture seems to be incompatible with the fact that
the superconducting transition temperature T$_{c}$ is closely
related to the number of CuO$_{2}$ planes per unit cell, with
single layer compounds of a family generically having lower T$_{c}$
than bilayer or trilayer compounds \cite{n2}. Additionally, there
are some subtle differences of the magnetic behaviors between doped
single layer and bilayer cuprates. By virtue of systematic studies
using NMR and $\mu$SR techniques, particularly the inelastic
neutron scattering, only incommensurate neutron scattering peaks
for the single layer lanthanum cuprate are observed in the
underdoped regime \cite{n5}, however, both low-energy
incommensurate neutron scattering peaks and high-energy
commensurate [$\pi$,$\pi$] resonance for the bilayer yttrium
cuprate in the normal state are detected \cite{n6}. These
experimental results highlight the importance of some sort of
coupling between the CuO$_{2}$ planes within a unit cell. It is
believed that all these experiments produce interesting data that
introduce the important constraints on the microscopic models and
theories.

The charge transport of doped single layer cuprates has been
addressed from several theoretical viewpoints \cite{n71,n72}.
Based on the charge-spin separation, an attractive proposal is
spinons and holons as basic low-energy excitations, serving as the
starting point for the gauge-theory approach \cite{n71}. It has
been shown \cite{n71} within the $t$-$J$ model that above the
Bose-Einstein temperature, the boson inverse lifetime due to
scattering by the gauge field is of order $T$, which suppresses
the condensation temperature and leads to a linear $T$ resistivity.
On the other hand, the spin-fermion model near the
antiferromagnetic instability has been developed to study the
normal-state properties of doped cuprates \cite{n72}. This
spin-fermion model describes low-energy fermions interacting with
their own collective spin fluctuations. Within this approach
\cite{n72}, the anomalous transport of doped single layer cuprates
has been studied extensively \cite{n72}, and the results are
consistent with the experiments.

As regards an intracell hopping, the band splitting in doped
bilayer cuprates was shown by the band calculation \cite{n7}, and
clearly observed \cite{n8,n9} recently by the
angle-resolved-photoemission spectroscopy in the doped bilayer
cuprate Bi$_{2}$Sr$_{2}$CaCu$_{2}$O$_{8+\delta}$ above T$_{c}$.
This bilayer band splitting is due to a nonvanishing intracell
coupling. Moreover, the magnitude of the bilayer splitting is
constant over a large range of dopings \cite{n9}. Considering these
highly unusual normal state properties in the underdoped regime
\cite{n1,n2,n5,n6}, a natural question is what is the effect of
the intracell coupling on the normal state properties of doped
bilayer cuprates. This is a challenge issue since the mechanism
for the superconductivity in doped cuprates has been widely
recognized to be closely related with the anisotropic normal-state
properties \cite{n10}. Based on the $t$-$J$ model, the charge
transport and spin response of doped single layer cuprates in the
underdoped regime have been discussed \cite{n11,n111,n12} within
the fermion-spin theory \cite{n13}, and the obtained results are
consistent with experiments \cite{n14}. In this paper, we apply
this successful approach to study the charge transport of the
underdoped bilayer cuprates. Our results show that although the
bilayer interaction leads to the band splitting in the electronic
structure, the qualitative behavior of the conductivity and
resistivity is the same as in the single layer case. The
conductivity shows the non-Drude behavior at low energies and
anomalous midinfrared band separated by the charge-transfer gap,
while the temperature dependent resistivity in the heavily
underdoped regime is characterized by a crossover from the high
temperature metallic-like to the low temperature insulating-like
behaviors.

We start from the bilayer $t$-$J$ model, which can be written as,
\begin{eqnarray}
H&=&-t\sum_{ai\hat{\eta}\sigma}C_{ai\sigma}^{\dagger}
C_{ai+\hat{\eta}\sigma}-t_{\perp}\sum_{i\sigma}
(C_{1i\sigma}^{\dagger}C_{2i\sigma}+{\rm h.c.}) \nonumber \\
&-&\mu\sum_{ai\sigma}C_{ai\sigma }^{\dagger }C_{ai\sigma }
+J\sum_{ai\hat{\eta}}{\bf S}_{ai}\cdot {\bf S}_{ai+\hat{\eta}}
\nonumber \\
&+&J_{\perp}\sum_i{\bf S}_{1i}\cdot {\bf S}_{2i},
\end{eqnarray}
where $\hat{\eta}=\pm\hat{x}$, $\pm\hat{y}$ within the plane, $a=1$
and $2$ is plane indices, $C^{\dagger}_{ai\sigma}$ ($C_{ai\sigma}$)
is the electron creation (annihilation) operator,
${\bf S}_{ai}=C_{ai}^{\dagger}{\vec{\sigma}}C_{ai}/2$ are spin
operators with ${\vec{\sigma}}=(\sigma_x,\sigma_y,\sigma_z)$ as
Pauli matrices, and $\mu$ is the chemical potential. The bilayer
$t$-$J$ model (1) is defined in the subspace with no doubly
occupied sites, {\it i.e.},
$\sum_{\sigma}C^{\dagger}_{ai\sigma}C_{ai\sigma}\leq 1$. The
strong electron correlation in the $t$-$J$ model manifests itself
by this single occupancy on-site local constraint \cite{n4}. To
deal with the local constraint in analytical calculations, the
fermion-spin theory \cite{n13},
$C_{ai\uparrow}=h^{\dagger}_{ai}S^{-}_{ai}$ and
$C_{ai\downarrow}=h^{\dagger}_{ai}S^{+}_{ai}$, has been proposed,
where the spinless fermion operator $h_{ai}$ keeps track of the
charge (holon), while the pseudospin operator $S_{ai}$ keeps track
of the spin (spinon), then it naturally incorporates the physics
of the charge-spin separation. In this case, the low-energy
behavior of the bilayer $t$-$J$ model (1) in the fermion-spin
representation can be rewritten as,
\begin{eqnarray}
H&=&t\sum_{ai\hat{\eta}}h^{\dagger}_{ai+\hat{\eta}}h_{ai}
(S^{+}_{ai}S^{-}_{ai+\hat{\eta}}+S^{-}_{ai}S^{+}_{ai+\hat{\eta}})
\nonumber \\
&+&t_{\perp}\sum_{i}(h^{\dagger}_{1i}h_{2i}+h^{\dagger}_{2i}
h_{1i})(S^{+}_{1i}S^{-}_{2i}+S^{-}_{1i}S^{+}_{2i})\nonumber \\
&+&\mu\sum_{ai}h^{\dagger}_{ai}h_{ai} + J_{{\rm eff}}
\sum_{ai\hat{\eta}}{\bf S}_{ai}\cdot {\bf S}_{ai+\hat{\eta}}
\nonumber \\
&+&J_{\perp{\rm eff}}\sum_i{\bf S}_{1i}\cdot {\bf S}_{2i},
\end{eqnarray}
with $J_{{\rm eff}}=J[(1-\delta)^{2}-\phi^{2}]$,
$J_{\perp {\rm eff}}=J_{\perp}[(1-\delta)^{2}-\phi^{2}_{\perp}]$,
the holon particle-hole order parameters
$\phi=\langle h^{\dagger}_{ai}h_{ai+\hat{\eta}}\rangle$ and
$\phi_{\perp}=\langle h^{\dagger}_{1i}h_{2i}\rangle$, $\delta$ is
the hole doping concentration, and $S^{+}_{ai}$ ($S^{-}_{ai}$)
is the pseudospin raising (lowering) operator. Since the single
occupancy local constraint has been treated properly within the
fermion-spin theory, this leads to disappearing of the extra gauge
degree of freedom related with this local constraint under the
charge-spin separation \cite{n13}. In this case, the charge
fluctuation couples only to holons \cite{n11,n111}. However, the
strong correlation between holons and spinons is still included
self-consistently through the spinon's order parameters entering
the holon's propagator, therefore both holons and spinons are
responsible for the charge transport. In this case, the
conductivity can be expressed as
$\sigma(\omega)=-{\rm Im}\Pi^{(h)}(\omega)/\omega$, with
$\Pi^{(h)}(\omega)$ is the holon current-current correlation
function, and is defined as $\Pi^{(h)}(\tau-\tau')=-\langle
T_{\tau}j^{(h)}(\tau)j^{(h)}(\tau')\rangle$, where $\tau$ and
$\tau'$ are the imaginary times, and $T_{\tau}$ is the $\tau$ order
operator. Within the Hamiltonian (2), the current density of holons
is obtained by the time derivation of the polarization operator
using Heisenberg's equation of motion as,
\begin{eqnarray}
j^{(h)}&=&2\chi et\sum_{ai\hat{\eta}}\hat{\eta}
h_{ai+\hat{\eta}}^{\dagger} h_{ai} \nonumber \\
&+&2\chi_{\perp}et_{\perp}\sum_{i}(R_{2i}-R_{1i})(h^{\dagger}_{2i}
h_{1i}-h^{\dagger}_{1i}h_{2i}),
\end{eqnarray}
where $R_{1i}$ ($R_{2i}$) is lattice site of the CuO$_{2}$ plane
$1$ (plane $2$),
$\chi=\langle S_{ai}^{+}S_{ai+\hat{\eta}}^{-}\rangle$ and
$\chi_{\perp}=\langle S^{+}_{1i}S^{-}_{2i}\rangle$ are the spinon
correlation functions, and $e$ is the electronic charge, which is
set as the unit hereafter. The holon current-current correlation
function can be calculated in terms of the holon Green's function
$g(k,\omega)$ as in the single layer case \cite{n11,n111}. However,
in the bilayer system, because there are two coupled CuO$_{2}$
planes, then the energy spectrum has two branches. In this case,
the one-particle holon Green's function can be expressed as a
matrix $g(i-j,\tau-\tau')=g_{L}(i-j,\tau-\tau')+\sigma_{x}
g_{T}(i-j,\tau-\tau')$, with the longitudinal and transverse parts
are defined as $g_{L}(i-j,\tau-\tau')=-\langle T_{\tau}h_{ai}(\tau)
h^{\dagger}_{aj}(\tau')\rangle$ and $g_{T}(i-j,\tau-\tau')=-\langle
T_{\tau}h_{ai}(\tau)h^{\dagger}_{a'j}(\tau')\rangle$ ($a\neq a'$),
respectively. Following discussions of the single layer case
\cite{n11,n111}, we obtain the conductivity of doped bilayer
cuprates as $\sigma(\omega)=\sigma^{(L)}(\omega)+\sigma^{(T)}
(\omega)$, with the longitudinal and transverse parts are given by,
\begin{mathletters}
\begin{eqnarray}
\sigma^{(L)}(\omega)&=&{1\over N}\sum_{k}[(2Z\chi t\gamma_{sk})^{2}
+(2\chi_{\perp} t_{\perp})^{2}]\times  \nonumber \\
&~&\int^{\infty}_{-\infty}{d\omega'\over 2\pi}A^{(h)}_{L}
(k,\omega'+\omega)A^{(h)}_{L}(k,\omega')\times \nonumber \\
&~&{n_{F}(\omega'+\omega)-n_{F}(\omega') \over \omega}, \\
\sigma^{(T)}(\omega)&=&{1\over N}\sum_{k}[(2Z\chi t\gamma_{sk})^{2}
-(2\chi_{\perp}t_{\perp})^{2}] \times \nonumber \\
&~&\int^{\infty}_{-\infty}{d\omega'\over 2\pi}A^{(h)}_{T}
(k,\omega'+\omega)A^{(h)}_{T}(k,\omega') \times \nonumber \\
&~&{n_{F}(\omega'+\omega)-n_{F}(\omega') \over \omega},
\end{eqnarray}
\end{mathletters}
respectively, where $Z$ is the coordination number within the
plane, $\gamma_{sk}=({\rm sin}k_{x}+{\rm sin}k_{y})/2$, and
$n_{F}(\omega)$ is the fermion distribution function. The
longitudinal and transverse holon spectral functions are obtained
as $A^{(h)}_{L}(k,\omega)=-2{\rm Im}g_{L}(k,\omega)$ and
$A^{(h)}_{T}(k,\omega)=-2{\rm Im} g_{T}(k,\omega)$, respectively.
The full holon Green's function $g^{-1}(k,\omega)=g^{(0)-1}
(k,\omega)-\Sigma^{(h)}(k,\omega)$ with the longitudinal and
transverse mean-field (MF) holon Green's functions,
$g^{(0)}_{L}(k,\omega)=1/2\sum_{\nu}1/(\omega-\xi^{(\nu)}_{k})$
and $g^{(0)}_{T}(k,\omega)=1/2\sum_{\nu}(-1)^{\nu+1}/(\omega-
\xi^{(\nu)}_{k})$, where $\nu=1$, $2$, and the longitudinal and
transverse second-order holon self-energy from the spinon pair
bubble are obtained by the loop expansion to the second-order as,
\begin{mathletters}
\begin{eqnarray}
\Sigma_{L}(k,\omega)&=&{1\over N^{2}}\sum_{pq}\sum_{\nu\nu'\nu''}
\Xi_{\nu\nu'\nu''}(k,p,q,\omega), \\
\Sigma_{T}(k,\omega)&=&{1\over N^{2}}\sum_{pq}\sum_{\nu\nu'\nu''}
(-1)^{\nu+\nu'+\nu''+1} \times \nonumber \\
&~&\Xi_{\nu\nu'\nu''}(k,p,q,\omega),
\end{eqnarray}
\end{mathletters}
respectively, with $\Xi_{\nu\nu'\nu''}(k,p,q,\omega)$ is given by,
\begin{eqnarray}
&~&\Xi_{\nu\nu'\nu''}(k,p,q,\omega)={B^{(\nu')}_{q+p}B^{(\nu)}_{q}
\over 32\omega^{(\nu')}_{q+p}\omega^{(\nu)}_{q}}\times \nonumber \\
&~&\left ( Zt[\gamma_{q+p+k}+\gamma_{q-k}]+t_{\perp}
[(-1)^{\nu+\nu''}+(-1)^{\nu'+\nu''}]\right )^{2}
\times \nonumber\\
&~&\left ({F^{(1)}_{\nu\nu'\nu''}(k,p,q)\over\omega+
\omega^{(\nu')}_{q+p}-\omega^{(\nu)}_{q}-\xi^{(\nu'')}_{p+k}}
\right. \left. +{F^{(2)}_{\nu\nu'\nu''}(k,p,q)\over\omega-
\omega^{(\nu')}_{q+p}+\omega^{(\nu)}_{q}-\xi^{(\nu'')}_{p+k}}
\right. \nonumber\\
&+&\left. {F^{(3)}_{\nu\nu'\nu''}(k,p,q)\over\omega+
\omega^{(\nu')}_{q+p}+\omega^{(\nu)}_{q}-\xi^{(\nu'')}_{p+k}}
\right. \nonumber \\
&+&\left. {F^{(4)}_{\nu\nu'\nu''}(k,p,q)\over\omega-
\omega^{(\nu')}_{q+p}-\omega^{(\nu)}_{q}-\xi^{(\nu'')}_{p+k}}
\right ),
\end{eqnarray}
with $\gamma_{k}=({\rm cos}k_x+{\rm cos}k_y)/2$, $B^{(\nu)}_{k}=
B_{k}-J_{\perp {\rm eff}}[\chi_{\perp}+2\chi^{z}_{\perp}(-1)^{\nu}]
[\epsilon_{\perp}+(-1)^{\nu}]$, $B_{k}=\lambda [(2\epsilon\chi^{z}+
\chi)\gamma_{k}-(\epsilon\chi+2\chi^{z})]$,
$\lambda=2ZJ_{\rm eff}$, $\epsilon=1+2t\phi/J_{\rm eff}$,
$\epsilon_{\perp}=1+4t_{\perp}\phi_{\perp}/J_{\perp{\rm eff}}$,
and
\begin{eqnarray}
F^{(1)}_{\nu\nu'\nu''}(k,p,q)&=&n_{F}(\xi^{(\nu'')}_{p+k})[n_{B}
(\omega^{(\nu)}_{q})-n_{B}(\omega^{(\nu')}_{q+p})] \nonumber \\
&+&n_{B}(\omega^{(\nu')}_{q+p})[1+n_{B}(\omega^{(\nu)}_{q})],
\nonumber \\
F^{(2)}_{\nu\nu'\nu''}(k,p,q)&=&n_{F}(\xi^{(\nu'')}_{p+k})[n_{B}
(\omega^{(\nu')}_{q+p})-n_{B}(\omega^{(\nu)}_{q})] \nonumber \\
&+&n_{B}(\omega^{(\nu)}_{q})[1+n_{B}(\omega^{(\nu')}_{q+p})],
\nonumber \\
F^{(3)}_{\nu\nu'\nu''}(k,p,q)&=&n_{F}(\xi^{(\nu'')}_{p+k})[1+n_{B}
(\omega^{(\nu')}_{q+p})+n_{B}(\omega^{(\nu)}_{q})] \nonumber \\
&+&n_{B}(\omega^{(\nu)}_{q})n_{B}(\omega^{(\nu')}_{q+p}),
\nonumber \\
F^{(4)}_{\nu\nu'\nu''}(k,p,q)&=&[1+n_{B}(\omega^{(\nu)}_{q})]
[1+n_{B}(\omega^{(\nu')}_{q+p})]\nonumber \\
&-&n_{F}(\xi^{(\nu'')}_{p+k})[1+n_{B}(\omega^{(\nu')}_{q+p})+n_{B}
(\omega^{(\nu)}_{q})],
\end{eqnarray}
where $n_{B}(\omega^{(\nu)}_{k})$ is the boson distribution
function, the MF holon excitation $\xi_{k}^{(\nu)}=2Z\chi
t\gamma_{k}+\mu+2\chi_{\perp}t_{\perp}(-1)^{\nu+1}$, the MF spinon
excitation $(\omega_{k}^{(\nu)})^{2}=\omega_{k}^{2}+\Delta_{k}^{2}
(-1)^{\nu+1}$, with $\omega_{k}^2=A_{1}\gamma_{k}^{2}+A_{2}
\gamma_{k}+A_{3}$, $\Delta_{k}^{2}=X_{1}\gamma_{k}+X_{2}$, and
\begin{eqnarray}
A_{1}&=&\alpha\epsilon\lambda^{2}(\chi/2+\epsilon\chi^{z}),
\nonumber\\
A_{2}&=&\epsilon\lambda^{2}[(1-Z)\alpha(\epsilon\chi/2+\chi^{z})
/Z \nonumber \\
&-&\alpha (C^z+C/2)-(1-\alpha)/(2Z)] \nonumber\\
&-&\alpha\lambda J_{\perp {\rm eff}}
[\epsilon (C_{\perp }^{z}+\chi_{\perp}^{z})+\epsilon_{\perp}
(C_{\perp}+\epsilon\chi_{\perp})/2], \nonumber\\
A_{3}&=&\lambda^{2}[\alpha (C^{z}+\epsilon^{2}C/2)+(1-\alpha)
(1+\epsilon^{2})/(4Z)\nonumber \\
&-&\alpha\epsilon (\chi/2+\epsilon\chi^{z})/Z]\nonumber\\
&+&\alpha\lambda J_{\perp {\rm eff}}[\epsilon\epsilon_{\perp}
C_{\perp}+2C_{\perp}^{z}]+J_{\perp {\rm eff}}^{2}
(\epsilon_{\perp}^{2}+1)/4, \nonumber\\
X_{1}&=&\alpha\lambda J_{\perp {\rm eff}}[(\epsilon_{\perp}\chi +
\epsilon\chi_{\perp})/2+\epsilon\epsilon_{\perp}(\chi_{\perp}^{z}+
\chi^{z})], \nonumber\\
X_{2}&=&-\alpha\lambda J_{\perp {\rm eff}}[\epsilon
\epsilon_{\perp}\chi/2+\epsilon_{\perp}(\chi^{z}+C_{\perp}^{z})+
\epsilon C_{\perp}/2]\nonumber\\
&-&\epsilon_{\perp}J_{\perp {\rm eff}}^{2}/2,
\end{eqnarray}
where the spinon correlation functions $\chi^{z}=\langle S_{ai}^{z}
S_{ai+\hat{\eta}}^{z}\rangle$, $\chi_{\perp}^{z}=\langle S_{1i}^{z}
S_{2i}^{z}\rangle$, $C=(1/Z^2)\sum_{\hat{\eta}\hat{\eta^{\prime}}}
\langle S_{ai+\hat{\eta}}^{+}S_{ai+\hat{\eta^{\prime}}}^{-}\rangle$,
and $C^{z}=(1/Z^2)\sum_{\hat{\eta}\hat{\eta ^{\prime }}}\langle
S_{ai+\hat{\eta}}^{z}S_{ai+\hat{\eta^{\prime }}}^{z}\rangle$,
$C_{\perp}=(1/Z)\sum_{\hat{\eta}}\langle S_{2i}^{+}
S_{1i+\hat{\eta}}^{-}\rangle$, and $C_{\perp}^{z}=(1/Z)
\sum_{\hat{\eta}}\langle S_{1i}^{z}S_{2i+\hat{\eta}}^{z}\rangle$.
In order to satisfy the sum rule for the correlation function
$\langle S_{ai}^{+}S_{ai}^{-}\rangle=1/2$ in the absence of the
antiferromagnetic long range order, a decoupling parameter
$\alpha$ has been introduced in the MF calculation, which can be
regarded as the vertex correction \cite{n15}. All these order
parameters, decoupling parameter $\alpha$, and the chemical
potential $\mu$ have been determined self-consistently, as done
in the single layer case \cite{n15}.

The frequency- and temperature-dependent conductivity is a powerful
probe for systems of interacting electrons, and provides very
detailed informations of the excitations, which interacts with
carriers in the normal-state and might play an important role in
the superconductivity. In Fig. 1, we present the results of the
conductivity $\sigma(\omega)$ at doping $\delta=0.05$ (solid line),
$\delta=0.06$ (dashed line), and $\delta=0.07$ (dotted line) for
parameters $t/J=2.5$, $t_{\perp}/t=0.25$, $J_{\perp}/J=0.25$ with
temperature $T=0$ in comparison with the experimental data
\cite{n16} taken on the underdoped YBa$_{2}$Cu$_{3}$O$_{7-x}$
(YBCO) (inset). The conductivity of bilayer cuprates in the
underdoped regime shows a sharp low-energy peak at $\omega<0.5t$
and the unusual midinfrared band appearing inside the
charge-transfer gap of the undoped system. After an  analysis, we
found that this low-energy peak decays fastly as
$\sigma(\omega)\sim 1/\omega$ (non-Drude fall-off) with increasing
energies. Moreover, the weight of the midinfrared peak is doping
dependent, and the peak position is shifted to low energy with
increasing dopings. For a better understanding of the optical
properties of doped bilayer cuprates, we have studied conductivity
at different temperatures, and the results at doping $\delta=0.06$
for $t/J=2.5$, $t_{\perp}/t=0.25$, and $J_{\perp}/J=0.25$ in $T=0$
(solid line), $T=0.3J$ (dashed line), and $T=0.5J$ (dotted line)
are plotted in Fig. 2 in comparison with the experimental data
\cite{n16} taken on the underdoped YBCO (inset). It is shown that
$\sigma(\omega)$ is temperature dependent, and the charge-transfer
gap is severely suppressed with increasing temperatures, and
vanishes at higher temperature ($T>0.4J$). Our results are in
qualitative agreement with the experiments \cite{n16}. In
comparison with the results from Refs. \cite{n11,n111}, it is shown
that the present conductivity also is qualitatively consistent with
these in the single layer case. In the above calculations, we also
find that the conductivity $\sigma(\omega)$ is essentially
determined by its longitudinal part $\sigma^{(L)}(\omega)$, this is
why in the present doped bilayer cuprates the conductivity spectrum
appears to reflect the single layer nature of the electronic state
\cite{n1,n2,n11,n111}. This is also why the in-plane charge
dynamics is rather universal within whole doped cuprates
\cite{n1,n2}.
\begin{figure}[prb]
\epsfxsize=2.5in\centerline{\epsffile{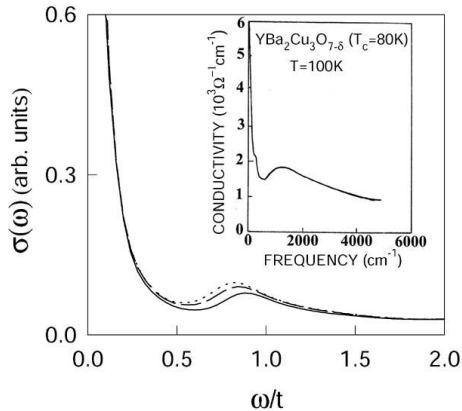}}
\caption{The conductivity at $\delta=0.05$ (solid line),
$\delta=0.06$ (dashed line), and $\delta=0.07$ (dotted line) for
$t/J=2.5$, $t_{\perp}/t=0.25$, and $J_{\perp}/J=0.25$ in the zero
temperature. Inset: the experimental result on the underdoped
YBa$_{2}$Cu$_{3}$O$_{7-x}$ taken from Ref. \cite{n16}.}
\end{figure}
\begin{figure}[prb]
\epsfxsize=2.5in\centerline{\epsffile{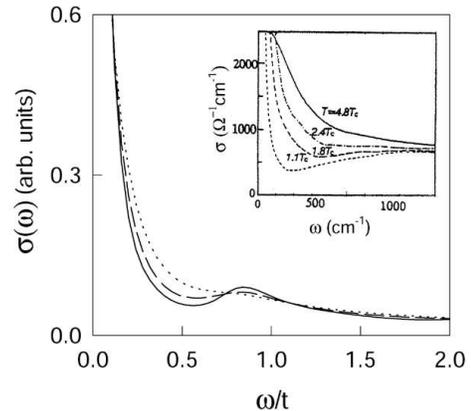}}
\caption{The conductivity at $\delta=0.06$ for $t/J=2.5$,
$t_{\perp}/t=0.25$, and $J_{\perp}/J=0.25$ in $T=0$ (solid line),
$T=0.3J$ (dashed line), and $T=0.5J$ (dotted line). Inset: the
experimental result on the underdoped YBa$_{2}$Cu$_{3}$O$_{7-x}$
taken from Ref. \cite{n16}.}
\end{figure}

Now we turn to discuss the resistivity, which is closely related
to the conductivity, and can be obtained as
$\rho(T)=1/\lim_{\omega\rightarrow 0}\sigma(\omega)$. This
resistivity has been calculated, and the results at doping
$\delta=0.05$ (solid line), $\delta=0.06$ (dashed line), and
$\delta=0.07$ (dotted line) for $t/J=2.5$, $t_{\perp}/t=0.25$,
and $J_{\perp}/J=0.25$ are plotted in Fig. 3 in comparison with
the experimental results \cite{n17} taken on the underdoped YBCO
(inset). These results show that in the heavily underdoped regime,
although the temperature-dependent resistivity is characterized
by a crossover from the high temperature metallic-like to the low
temperature insulating-like behaviors, the nearly temperature liner
dependence in the resistivity dominates over a wide temperature
range, in agreement with the experimental results \cite{n17}. In
comparison with the results from Refs. \cite{n11,n111}, it is shown
that the present resistivity also is qualitatively consistent with
these in the single layer case. We emphasize that since the order
parameters, decoupling parameter $\alpha$, and the chemical
potential $\mu$ have been determined self-consistently, then these
theoretical results were obtained without any adjustable
parameters. Furthermore, it is found in the above discussions that
the present results are insensitive to the reasonable values of
$t/J$, $t_{\perp}/t$, and $J_{\perp}/J$ as in the single layer
case \cite{n11,n111}.
\begin{figure}[prb]
\epsfxsize=2.5in\centerline{\epsffile{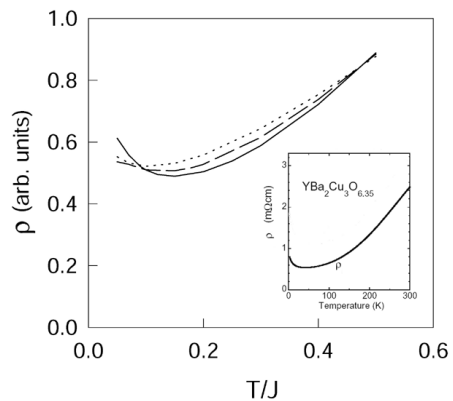}}
\caption{
The resistivity at $\delta=0.05$ (solid line),
$\delta=0.06$ (dashed line), and $\delta=0.07$ (dotted line) for
$t/J=2.5$, $t_{\perp}/t=0.25$, and $J_{\perp}/J=0.25$. Inset: the
experimental result on the underdoped YBa$_{2}$Cu$_{3}$O$_{7-x}$
taken from Ref. \cite{n17}.}
\end{figure}

An explanation for the metal-to-insulating crossover in the
resistivity in the heavily underdoped regime can be found from the
competition between the kinetic energy and magnetic energy in the
system. Since cuprate superconducting materials are doped Mott
insulators, obtained by chemically adding charge carriers to a
strongly correlated antiferromagnetic insulating state, therefore
doped cuprates are characterized by the competition between the
kinetic energy ($t$) and magnetic energy ($J$). The magnetic
energy $J$ favors the magnetic order for spins, while the kinetic
energy $t$ favors delocalization of holes and tends to destroy the
magnetic order. In the present fermion-spin theory, although both
holons and spinons contribute to the charge transport, the
scattering of holons dominates the charge transport \cite{n11},
where the charged holon scattering rate is obtained from the full
holon Green's function (then the holon self-energy (5) and holon
spectral function) by considering the holon-spinon interaction,
therefore in the heavily underdoped regime the observed crossover
from the high temperature metallic-like to the low temperature
insulating-like behaviors in the resistivity is closely related
with this competition. In lower temperatures, the holon kinetic
energy is much smaller than the magnetic energy, in this case the
magnetic fluctuation is strong enough to severely reduce the
charged holon scattering and thus is responsible for the
insulating-like behavior in the resistivity. With increasing
temperatures, the holon kinetic energy is increased, while the
spinon magnetic energy is decreased. In the region where the holon
kinetic energy is much larger than the spinon magnetic energy at
higher temperatures, the charged holon scattering would give rise
to the temperature linear resistivity.

In summary, we have studied the charge transport in the underdoped
bilayer cuprates by considering the bilayer interaction. It is
shown that although the bilayer interaction leads to the band
splitting in the electronic structure, the qualitative behavior of
the charge transport is the same as in the single layer case. The
conductivity spectrum shows a low-energy peak and the anomalous
midinfrared band. This midinfrared band is suppressed severely
with increasing temperatures, while the resistivity exhibits a
crossover from the high temperature metallic-like to the low
temperature insulating-like behaviors. Our results also show that
the mechanism that cause this unusual charge transport in the
underdoped cuprates is closely related to the background
antiferromagnetic correlations.

\acknowledgments
This work was supported by the National Natural Science Foundation
of China under Grant Nos. 10074007, 10125415, and 90103024, the
special funds from the Ministry of Science and Technology of China,
and the National Science Council under Grant
No. NSC 90-2816-M-032-0001-6.

\end{document}